\documentclass[sigconf]{acmart}

\usepackage{fancyvrb}
\usepackage{xurl}
\usepackage{booktabs}
\usepackage{url}
\usepackage{amsmath,amsfonts}
\usepackage{algorithmic}
\usepackage{graphicx}
\usepackage{textcomp}
\usepackage{xspace}
\usepackage{subcaption}
\usepackage{tabularx}
\usepackage{framed}
\usepackage{chngpage}
\usepackage{pgfplots}
\pgfplotsset{compat=newest}
\usepackage{siunitx}
\usepackage{balance}
\usepackage{enumitem}
\usepackage{listings} 
\usepackage{multirow} 
\usepackage{xfp}      
\usepackage{url}

\definecolor{awesome}{RGB}{0,0,0} 

\usepackage{hyperref}
\usepackage{makecell}

\usepackage{tikz}

\newcommand{\rqa}{$RQ_1$}
\newcommand{\rqb}{$RQ_2$}
\newcommand{\rqc}{$RQ_3$}

\newcommand{\rqaa}{How often do developers add tests with Agent Coding Tools?}
\newcommand{\rqbb}{What is the difference between AI- and human-created test methods?}
\newcommand{\rqcc}{Do Agent Coding Tools improve test coverage?}

\newcommand{\rqA}{\rqa: \rqaa}
\newcommand{\rqB}{\rqb: \rqbb}
\newcommand{\rqC}{\rqc: \rqcc}

\newcommand{\TotalFuzzingProjectNum}{1311}
\newcommand{\TargetProjectNum}{628}

\pgfmathsetmacro{\TargetProjectRateNum}{round((\TargetProjectNum/\TotalFuzzingProjectNum)*100*10)/10}

\newcommand{\ExcludedProjectNum}{4}
\pgfmathsetmacro{\FinallyTargetProjectInt}{round(\TargetProjectNum - \ExcludedProjectNum)}

\definecolor{darkgreen}{rgb}{0, 0.5, 0} 
\definecolor{whitesmoke}{rgb}{0.99, 0.99, 0.99} 

\def\Underline{\setbox0\hbox\bgroup\let\\\endUnderline}
\def\endUnderline{\vphantom{y}\egroup\smash{\underline{\box0}}\\}
\def\|{\verb|}

\newcommand{\eg}{\textit{e.g.,}\xspace}

\newcounter{findings_no}

\usepackage{listings}
\definecolor{backcolour}{rgb}{0.95,0.95,0.92}
\lstdefinelanguage{diff}{
  morecomment=**[f][\color{red}]{-},         
  morecomment=**[f][\color{darkgreen}]{+},       
  moredelim=**[is][\bfseries]{@@}{@@},
}
\definecolor{backcolour}{rgb}{0.95,0.95,0.92}
\lstdefinelanguage{commit}{ 
  breakindent = 0pt,
  numbers=none,
  backgroundcolor=\color{white},
  frame=single,
  xleftmargin=3.5em,
  numbersep=0em,
  xrightmargin=1.5em,
}

\definecolor{main}{HTML}{D0D3D4}    
\definecolor{sub}{HTML}{D0D3D4}     

\copyrightyear{2026}
\acmYear{2026}
\setcopyright{cc}
\setcctype{by}
\acmConference[MSR ’26]{23rd International Conference on Mining Software Repositories}{April 13--14, 2026}{Rio de Janeiro, Brazil}
\acmBooktitle{23rd International Conference on Mining Software Repositories (MSR ’26), April 13--14, 2026, Rio de Janeiro, Brazil}
\acmPrice{}
\acmDOI{10.1145/3793302.3793620}
\acmISBN{979-8-4007-2474-9/2026/04}

\begin{document}

\DeclareRobustCommand{\calcpct}[2]{\fpeval{round((#1/#2)*100, 2)}} 

\AtBeginDocument{%
  \providecommand\BibTeX{{%
    \normalfont B\kern-0.5em{\scshape i\kern-0.25em b}\kern-0.8em\TeX}}}

\title{Testing with AI Agents: An Empirical Study of Test Generation Frequency, Quality, and Coverage}







\author{
  Suzuka Yoshimoto\textsuperscript{$\dagger$},
  Shun Fujita\textsuperscript{$\dagger$},
  Kosei Horikawa\textsuperscript{$\dagger$},
  Daniel Feitosa\textsuperscript{$\S$},
  Yutaro Kashiwa\textsuperscript{$\dagger$},
  Hajimu Iida\textsuperscript{$\dagger$}
}

\affiliation{%
  \institution{\textsuperscript{$\dagger$}Nara Institute of Science and Technology, Japan}
  \country{}
}
\affiliation{%
  \institution{\textsuperscript{$\S$}University of Groningen, the Netherlands}
  \country{}
}

\email{{yoshimoto.suzuka.yw4, fujita.shun.fp9, horikawa.kosei.hk1, yutaro.kashiwa, iida}@naist.ac.jp}
\email{d.feitosa@rug.nl}

\renewcommand{\shortauthors}{Yoshimoto, et al.}


\newcommand{\liamTotalCommits}{6745\xspace}
\newcommand{\liamTestAddCommits}{352\xspace}
\newcommand{\liamTestAddRatio}{5.2\%\xspace}
\newcommand{\liamAiTestAddCommits}{43\xspace}
\newcommand{\liamAiTestAddRatio}{12.2\%\xspace}

\newcommand{\calcomTotalCommits}{21840\xspace}
\newcommand{\calcomTestAddCommits}{894\xspace}
\newcommand{\calcomTestAddRatio}{4.1\%\xspace}
\newcommand{\calcomAiTestAddCommits}{129\xspace}
\newcommand{\calcomAiTestAddRatio}{14.4\%\xspace}

\newcommand{\appkitTotalCommits}{3857\xspace}
\newcommand{\appkitTestAddCommits}{257\xspace}
\newcommand{\appkitTestAddRatio}{6.7\%\xspace}
\newcommand{\appkitAiTestAddCommits}{29\xspace}
\newcommand{\appkitAiTestAddRatio}{11.3\%\xspace}

\newcommand{\capgoTotalCommits}{274\xspace}
\newcommand{\capgoTestAddCommits}{24\xspace}
\newcommand{\capgoTestAddRatio}{8.8\%\xspace}
\newcommand{\capgoAiTestAddCommits}{20\xspace}
\newcommand{\capgoAiTestAddRatio}{83.3\%\xspace}

\newcommand{\totalTSTotalCommits}{94\xspace}
\newcommand{\totalTSTestAddCommits}{26\xspace}
\newcommand{\totalTSTestAddRatio}{27.7\%\xspace}
\newcommand{\totalTSAiTestAddCommits}{26\xspace}
\newcommand{\totalTSAiTestAddRatio}{100.0\%\xspace}

\newcommand{\primerReactTotalCommits}{1342\xspace}
\newcommand{\primerReactTestAddCommits}{89\xspace}
\newcommand{\primerReactTestAddRatio}{6.6\%\xspace}
\newcommand{\primerReactAiTestAddCommits}{31\xspace}
\newcommand{\primerReactAiTestAddRatio}{34.8\%\xspace}

\newcommand{\alchemyTotalCommits}{679\xspace}
\newcommand{\alchemyTestAddCommits}{76\xspace}
\newcommand{\alchemyTestAddRatio}{11.2\%\xspace}
\newcommand{\alchemyAiTestAddCommits}{54\xspace}
\newcommand{\alchemyAiTestAddRatio}{71.1\%\xspace}

\newcommand{\llmgatewayTotalCommits}{1154\xspace}
\newcommand{\llmgatewayTestAddCommits}{25\xspace}
\newcommand{\llmgatewayTestAddRatio}{2.2\%\xspace}
\newcommand{\llmgatewayAiTestAddCommits}{14\xspace}
\newcommand{\llmgatewayAiTestAddRatio}{56.0\%\xspace}

\newcommand{\helperTotalCommits}{3688\xspace}
\newcommand{\helperTestAddCommits}{123\xspace}
\newcommand{\helperTestAddRatio}{3.3\%\xspace}
\newcommand{\helperAiTestAddCommits}{13\xspace}
\newcommand{\helperAiTestAddRatio}{10.6\%\xspace}

\newcommand{\azureSdkTotalCommits}{3864\xspace}
\newcommand{\azureSdkTestAddCommits}{366\xspace}
\newcommand{\azureSdkTestAddRatio}{9.5\%\xspace}
\newcommand{\azureSdkAiTestAddCommits}{7\xspace}
\newcommand{\azureSdkAiTestAddRatio}{1.9\%\xspace}

\newcommand{\valLiamTestAdd}{352}
\newcommand{\valLiamAiTestAdd}{43}

\newcommand{\valCalcomTestAdd}{894}
\newcommand{\valCalcomAiTestAdd}{129}

\newcommand{\valAppkitTestAdd}{257}
\newcommand{\valAppkitAiTestAdd}{29}

\newcommand{\valCapgoTestAdd}{24}
\newcommand{\valCapgoAiTestAdd}{20}

\newcommand{\valTotalTSTestAdd}{26}
\newcommand{\valTotalTSAiTestAdd}{26}

\newcommand{\valPrimerReactTestAdd}{89}
\newcommand{\valPrimerReactAiTestAdd}{31}

\newcommand{\valAlchemyTestAdd}{76}
\newcommand{\valAlchemyAiTestAdd}{54}

\newcommand{\valLlmTestAdd}{25}
\newcommand{\valLlmAiTestAdd}{14}

\newcommand{\valHelperTestAdd}{123}
\newcommand{\valHelperAiTestAdd}{13}

\newcommand{\valAzureTestAdd}{366}
\newcommand{\valAzureAiTestAdd}{7}

\newcommand{\valLiamTotal}{6745}
\newcommand{\valCalcomTotal}{21840}
\newcommand{\valAppkitTotal}{3857}
\newcommand{\valCapgoTotal}{274}
\newcommand{\valTotalTSTotal}{94}
\newcommand{\valPrimerReactTotal}{1342}
\newcommand{\valAlchemyTotal}{679}
\newcommand{\valLlmTotalProject}{1154}
\newcommand{\valHelperTotal}{3688}
\newcommand{\valAzureTotal}{3864}

\newcommand{\valTotalAnalyzedTestCommits}{%
    \inteval{%
        \valLiamTestAdd + \valCalcomTestAdd + \valAppkitTestAdd + 
        \valCapgoTestAdd + \valTotalTSTestAdd + \valPrimerReactTestAdd + 
        \valAlchemyTestAdd + \valLlmTestAdd + \valHelperTestAdd + \valAzureTestAdd
    }%
}

\newcommand{\valTotalAiTestAddCommitsOnly}{%
    \inteval{%
        \valLiamAiTestAdd + \valCalcomAiTestAdd + \valAppkitAiTestAdd + 
        \valCapgoAiTestAdd + \valTotalTSAiTestAdd + \valPrimerReactAiTestAdd + 
        \valAlchemyAiTestAdd + \valLlmAiTestAdd + \valHelperAiTestAdd + \valAzureAiTestAdd
    }%
}

\newcommand{\valTotalProjectCommitsOnly}{%
    \inteval{%
        \valLiamTotal + \valCalcomTotal + \valAppkitTotal + 
        \valCapgoTotal + \valTotalTSTotal + \valPrimerReactTotal + 
        \valAlchemyTotal + \valLlmTotalProject + \valHelperTotal + \valAzureTotal
    }%
}

\newcommand{\totalAnalyzedCommits}{\num{\valTotalAnalyzedTestCommits}\xspace}

\newcommand{\totalAiRatioAmongTests}{%
    \fpeval{round(\valTotalAiTestAddCommitsOnly / \valTotalAnalyzedTestCommits * 100, 2)}\%\xspace
}

\newcommand{\totalTestAddRatio}{%
    \fpeval{round(\valTotalAnalyzedTestCommits / \valTotalProjectCommitsOnly * 100, 2)}\%\xspace
}

\begin{abstract}
Agent-based coding tools have transformed software development practices. Unlike prompt-based approaches that require developers to manually integrate generated code, these agent-based tools autonomously interact with repositories to create, modify, and execute code, including test generation. While many developers have adopted agent-based coding tools, little is known about how these tools generate tests in real-world development scenarios or how AI-generated tests compare to human-written ones.

This study presents an empirical analysis of test generation by agent-based coding tools using the AIDev dataset. We extracted \totalAnalyzedCommits commits containing test-related changes and investigated three aspects: the frequency of test additions, the structural characteristics of the generated tests, and their impact on code coverage. Our findings reveal that (i) AI authored \totalAiRatioAmongTests of all commits adding tests in real-world repositories, (ii) AI-generated test methods exhibit distinct structural patterns, featuring longer code and a higher density of assertions while maintaining lower cyclomatic complexity through linear logic, and (iii) AI-generated tests contribute to code coverage comparable to human-written tests, frequently achieving positive coverage gains across several projects.

\end{abstract}



\maketitle

\section{Introduction}\label{sec:introduction}
Software testing is an essential activity that forms the foundation of quality assurance~\cite{DBLP:books/daglib/0019907, DBLP:conf/icse/KoDD14, DBLP:conf/scam/KhatamiZ23}. Code lacking adequate tests cannot only delay bug detection but also increase the risk of regressions and undermine developers' confidence in making changes~\cite{DBLP:books/oreilly/WintersMS20, DBLP:conf/icst/LeinenEPSSJ24}. Furthermore, low test coverage can discourage code modifications and extensions, thereby hindering the evolution of projects \cite{DBLP:journals/tse/KimZN14, DBLP:conf/icsm/SpadiniPZBB18}. To avoid such problems, developers are encouraged to write tests in parallel with feature implementation; however, due to time constraints and the tedious nature of test creation, sufficient tests are often not written in practice \cite{DBLP:conf/sbes/SouzaFRRSM20, DBLP:conf/icse/BellerGZ15}. To address this, automated test generation has evolved from traditional techniques like symbolic execution~\cite{DBLP:journals/csur/BaldoniCDDF18} to recent approaches leveraging Large Language Models (LLMs)~\cite{DBLP:journals/tosem/HouZLYWLLLGW24}.

More recently, the mainstream approach has been shifting from single-prompt-based methods to approaches utilizing AI agents. In agent-based approaches, by understanding the context of the entire repository and autonomously repeating cycles of test creation, execution, and modification, more sophisticated test generation becomes possible~\cite{10923987}. Although existing studies have evaluated the quality and coverage of LLM-generated tests in controlled experimental environments~\cite{DBLP:journals/tse/SchaferNET24}, it is not sufficiently understood to what extent agent-based tools generate tests in actual development settings and how the generated tests differ from those created by humans. This represents an important research challenge for evaluating the practicality of agent-based test generation technology.

This study aims to quantitatively analyze the reality of test generation by agent-based coding tools. Specifically, we extracted commits containing test-related changes from the AIDev dataset \cite{DBLP:journals/corr/abs-2507-15003} and investigated the frequency of test generation, characteristics, and impact on coverage in those commits. We identified \totalAnalyzedCommits commits and compared test methods generated by AI with those created by humans, thereby conducting a multifaceted analysis of the effects of test generation by agent-based tools. 


\newcommand{\liamPRs}{34\xspace}
\newcommand{\liamRelease}{@liam-hq/erd-core@0.6.15\xspace}
\newcommand{\calcomPRs}{33\xspace}
\newcommand{\calcomRelease}{v5.8.12\xspace}
\newcommand{\appkitPRs}{21\xspace}
\newcommand{\appkitRelease}{@reown/appkit@1.8.11\xspace}
\newcommand{\capgoPRs}{16\xspace}
\newcommand{\capgoRelease}{v12.49.4\xspace}
\newcommand{\totalTSPRs}{16\xspace}
\newcommand{\totalTSRelease}{@total-typescript/\\twoslash-to-simple-markdown@0.2.0\xspace}
\newcommand{\primerReactPRs}{16\xspace}
\newcommand{\primerReactRelease}{@primer/react@38.1.0\xspace}
\newcommand{\alchemyPRs}{14\xspace}
\newcommand{\alchemyRelease}{v0.76.1\xspace}
\newcommand{\llmgatewayPRs}{10\xspace}
\newcommand{\llmgatewayRelease}{v1.2.0\xspace}
\newcommand{\helperPRs}{9\xspace}
\newcommand{\helperRelease}{v0.5.0\xspace}
\newcommand{\azureSdkPRs}{7\xspace}
\newcommand{\azureSdkRelease}{@azure/arm-deviceregistry\_1.1.0\xspace}
\section{Data Collection}\label{sec:data_procedure}
To identify projects suitable for our analysis, we applied the following steps using the AIDev dataset~\cite{DBLP:journals/corr/abs-2507-15003}.

\smallskip
\noindent\textbf{1. Language Filtering}: We analyzed the AIDev dataset to identify the languages with the most AI-authored pull requests (PRs). We found TypeScript to be one of the most common languages across all agents (22.4\% of PRs) and, thus, selected it for this study.

\smallskip\noindent\textbf{2. Framework Filtering}: To measure coverage systematically (for RQ3), we decided to standardize the test framework use to make this study feasible. In particular, we investigated the 2,807 unique repositories from the AIDev dataset that contain PRs, finding that 650 were identified as TypeScript projects, of which 400 had detectable test frameworks. Among these, \emph{vitest} was found to be the most prevalent, used in 179 projects (44.8\% of projects with test frameworks). Given \emph{vitest}'s dominance in the TypeScript ecosystem within our dataset and its straightforward coverage measurement capabilities, we decided to target \emph{vitest} projects for our analysis.


\smallskip
\noindent\textbf{3. Project Identification}: Next, focusing on the identified \emph{vitest} projects, we returned to the AIDev dataset to find those with a high volume of AI-authored test additions. We inspected the \texttt{patch} information of commits within AI-authored PRs. We filtered for PRs that contained new test method definitions (e.g., \texttt{describe()}, \texttt{it()}) on added lines (prefixed with \texttt{+}).


\smallskip
\noindent\textbf{Final Dataset:} We selected the top 10 projects with the highest AI-authored test activity to ensure a focus on the most significant contributions. This sample size was determined by the technical complexity of dynamic analysis; measuring test coverage (RQ3) requires manual environment setup and dependency resolution for each repository, making a larger-scale execution impractical. 
\makeatletter
\newsavebox{\acmbox@box}
\newlength{\acmbox@innerwd}
\newenvironment{acmbox}[1]{%
  \def\acmbox@title{#1}%
  \par\medskip\noindent
  \setlength{\fboxsep}{6pt}%
  \setlength{\fboxrule}{0.8pt}%
  \setlength{\acmbox@innerwd}{\dimexpr\linewidth-2\fboxsep-2\fboxrule\relax}%
  \begin{lrbox}{\acmbox@box}%
  \begin{minipage}{\acmbox@innerwd}%
  \setlength{\parindent}{0pt}%
  \setlength{\parskip}{0.35\baselineskip}%
  {\color{awesome}\bfseries \acmbox@title}\par\vspace{2pt}%
}{%
  \end{minipage}%
  \end{lrbox}%
  \noindent\fcolorbox{awesome}{gray!5}{\usebox{\acmbox@box}}%
  \par\medskip
}
\makeatother

\setlength{\textfloatsep}{5pt}
\begin{table}[t]
    \centering
    \small
    \setlength{\abovecaptionskip}{5pt}
    \setlength{\belowcaptionskip}{2pt}
    \caption{Prevalence of AI-Generated Test Addition Commits in Selected Projects.}
    \label{tab:rq1_results}
    \begin{tabular}{l r r}
        \toprule
        \multirow{2}{*}{\textbf{Project}} & \multicolumn{2}{c}{\textbf{Test-adding Commits}} \\
        \cmidrule(lr){2-3}
        & \textbf{All} & \textbf{AI-authored (\%)} \\
        \midrule
        cal.com~\cite{repo:calcom} & \calcomTestAddCommits & \calcomAiTestAddCommits~(\calcomAiTestAddRatio) \\
        azure-sdk-for-js~\cite{repo:azure-sdk} & \azureSdkTestAddCommits & \azureSdkAiTestAddCommits~(\azureSdkAiTestAddRatio) \\
        liam~\cite{repo:liam} & \liamTestAddCommits & \liamAiTestAddCommits~(\liamAiTestAddRatio) \\
        appkit~\cite{repo:appkit} & \appkitTestAddCommits & \appkitAiTestAddCommits~(\appkitAiTestAddRatio) \\
        helper~\cite{repo:helper} & \helperTestAddCommits & \helperAiTestAddCommits~(\helperAiTestAddRatio) \\
        primer-react~\cite{repo:primer-react} & \primerReactTestAddCommits & \primerReactAiTestAddCommits~(\primerReactAiTestAddRatio) \\
        alchemy~\cite{repo:alchemy} & \alchemyTestAddCommits & \alchemyAiTestAddCommits~(\alchemyAiTestAddRatio) \\
        \shortstack[l]{total-typescript-monorepo}~\cite{repo:total-typescript} & \totalTSTestAddCommits & \totalTSAiTestAddCommits~(\totalTSAiTestAddRatio) \\ 
        llmgateway~\cite{repo:llmgateway} & \llmgatewayTestAddCommits & \llmgatewayAiTestAddCommits~(\llmgatewayAiTestAddRatio) \\
        capgo~\cite{repo:capgo} & \capgoTestAddCommits & \capgoAiTestAddCommits~(\capgoAiTestAddRatio) \\ 
        \bottomrule
    \end{tabular}%
\end{table}

\section{Research Questions}\label{sec:results}
\subsection*{\rqA}\label{sec:rqa}
\noindent\textbf{Motivation.} Understanding the prevalence of AI-generated test contributions is the first step in assessing their overall impact. We examine the proportion of test-added commits among all commits made by AI agents to gauge their activity level in this specific task.

\smallskip\noindent\textbf{Approach.} 
We analyzed 10 selected TypeScript projects to quantify the contribution of AI agents to test code generation. Our methodology focuses on identifying commits where AI agents explicitly introduced new test methods. The process consists of four steps:


\smallskip
\noindent\textit{1. Defining the Temporal Scope.}
We determined the analysis period for each project based on the available data in the AIDev dataset. By examining the \texttt{pull\_request} table, we extracted the minimum and maximum \texttt{created\_at} timestamps for each repository. This defined interval served as the boundary for our data collection, ensuring that our commit analysis corresponds precisely to the period of AI activity captured in the dataset.

\smallskip
\noindent\textit{2. Commit Retrieval and Filtering.}
We utilized the GitHub API to retrieve all commits associated with pull requests within the defined period. To isolate substantive code contributions, we filtered this set to exclude merge commits. Furthermore, we narrowed our focus to commits that modified at least one test file, as these are the potential candidates for test additions.

\smallskip
\noindent\textit{3. Detecting Test Method Additions.}
Modification of a test file does not necessarily guarantee that a new test was added; it could be a refactor or a deletion. To accurately identify test additions, we performed a granular analysis of the commit \texttt{patch}. We inspected only the added lines (prefixed with \texttt{+}) and applied regular expressions to detect the syntax of new test definitions common in the \emph{vitest} and TypeScript ecosystem. Specifically, we looked for patterns such as \texttt{test()}, \texttt{it()}, \texttt{describe()}, \texttt{test.skip()}, \texttt{test.only()}, \texttt{xtest}, and \texttt{ftest}. A commit was classified as a ``Test Addition Commit'' only if its patch contained these specific markers of new test cases.

\smallskip
\noindent\textit{4. AI Author Identification.}
\label{step:author-id}
We determined the authorship of the identified Test Addition Commits by checking the committer name to determine whether the author was an AI or a human. We used AST parsing to extract test methods and \texttt{git blame} to trace the authorship of each line within those methods. A commit was classified as AI-authored only if all lines within all of its added test methods were authored exclusively by AI agents. 

\smallskip
\noindent\textit{5. Metrics calculation.}
We computed the proportion of test-adding commits attributed to AI contributors.

\smallskip
\noindent\textbf{Results.}
Table~\ref{tab:rq1_results} summarizes the prevalence of AI-authored test additions across the selected projects.
The results reveal a highly polarized landscape regarding the role of AI in test generation.

\smallskip
\noindent\textbf{Shift in AI Role by Project Scale.}
We observe that the role of AI agents in test generation significantly varies with the project's scale, which we characterize by the number of human contributors. In small-scale or early-stage projects with a limited contributor base, AI agents often take a \textit{dominant} role. For instance, in \texttt{total-typescript-monorepo} (5 contributors) and \texttt{capgo} (28 contributors), AI-generated tests account for 100\% and 83.3\% of test additions, respectively. 
In contrast, for large-scale projects with hundreds of contributors, the AI contribution shifts to an \textit{auxiliary} role. In the largest analyzed project, \texttt{cal.com} (876 contributors), the AI contribution is consistent but remains at \calcomAiTestAddRatio. This trend is even more pronounced in established enterprise projects like \texttt{azure-sdk-for-js} (608 contributors), where the AI contribution drops to a marginal \azureSdkAiTestAddRatio. These results suggest that while AI agents can drive testing in small teams, they mainly serve as supportive tools in large-scale, community-driven, or enterprise development.

\smallskip
\noindent\textbf{High AI Dominance in Specific Projects.}
In three projects, AI agents are the primary or exclusive drivers of test creation. In \texttt{total-typescript-monorepo}, 100\% of detected test additions were authored by AI. Similarly, \texttt{capgo} (\capgoAiTestAddRatio) and \texttt{alchemy} (\alchemyAiTestAddRatio) show very high AI participation. This suggests that in certain contexts (possibly smaller, experimental, or agent-centric projects), AI tools can already take over testing responsibility, or developers rely on them enough to delegate the task.



\begin{acmbox}

    \textbf{Answer to RQ1.}
    AI contribution to test generation varies by project type: up to 100\% in agent-heavy projects, but typically 10\% to 15\% in broader open-source development.
    
\end{acmbox}



\subsection*{\rqB}\label{sec:rqb}
\noindent\textbf{Motivation.}
Understanding whether AI-generated tests differ from human-written ones in structure and content is important, as this impacts their quality and maintainability~\cite{xUnitTestPatterns}. 

\noindent\textbf{Approach.}
We cloned the repositories of the 10 selected projects and checked out the latest release for each, ensuring our analysis was performed on a consistent, stable snapshot of the codebase.
We then identified the authorship of individual test methods following the line-level attribution methodology in Step 4 of the RQ1 Approach (AI Author Identification). Based on this, we classified methods as either AI Test Method (all lines authored by AI) or Human Test Method (any line authored by a human). Finally, we compared the two datasets (AI Test Methods and Human Test Methods) using two techniques. 

\textit{1. Metric-Based Analysis:} We calculated metrics for each test method: Effective Lines of Code (\texttt{eLoC}, excluding comments and empty lines), number of assertions (\texttt{\# assertions}, e.g., \texttt{expect()}), and Cyclomatic Complexity (\texttt{CC}).
\texttt{eLoC} and \texttt{\# assertions} were calculated using regular expressions. \texttt{CC} was calculated by parsing each test method into an AST using Babel (\texttt{@babel/parser}) and counting decision-point nodes (e.g., \texttt{If}, \texttt{For}) via \texttt{@babel/traverse}.

\textit{2. Embedding Visualization:} We generated vector embeddings for each test method using a pre-trained CodeBERT model to capture deeper semantic and structural characteristics. We then used t-SNE (t-distributed stochastic neighbor embedding) to visualize these high-dimensional embeddings in 2D space, allowing us to observe whether AI- and human-generated tests form distinct clusters.

\begin{figure}[t]
    \centering
    \captionsetup[subfigure]{aboveskip=2pt, belowskip=5pt}
    
    \begin{subfigure}[b]{0.46\textwidth}
        \centering
        \includegraphics[width=0.8\linewidth]{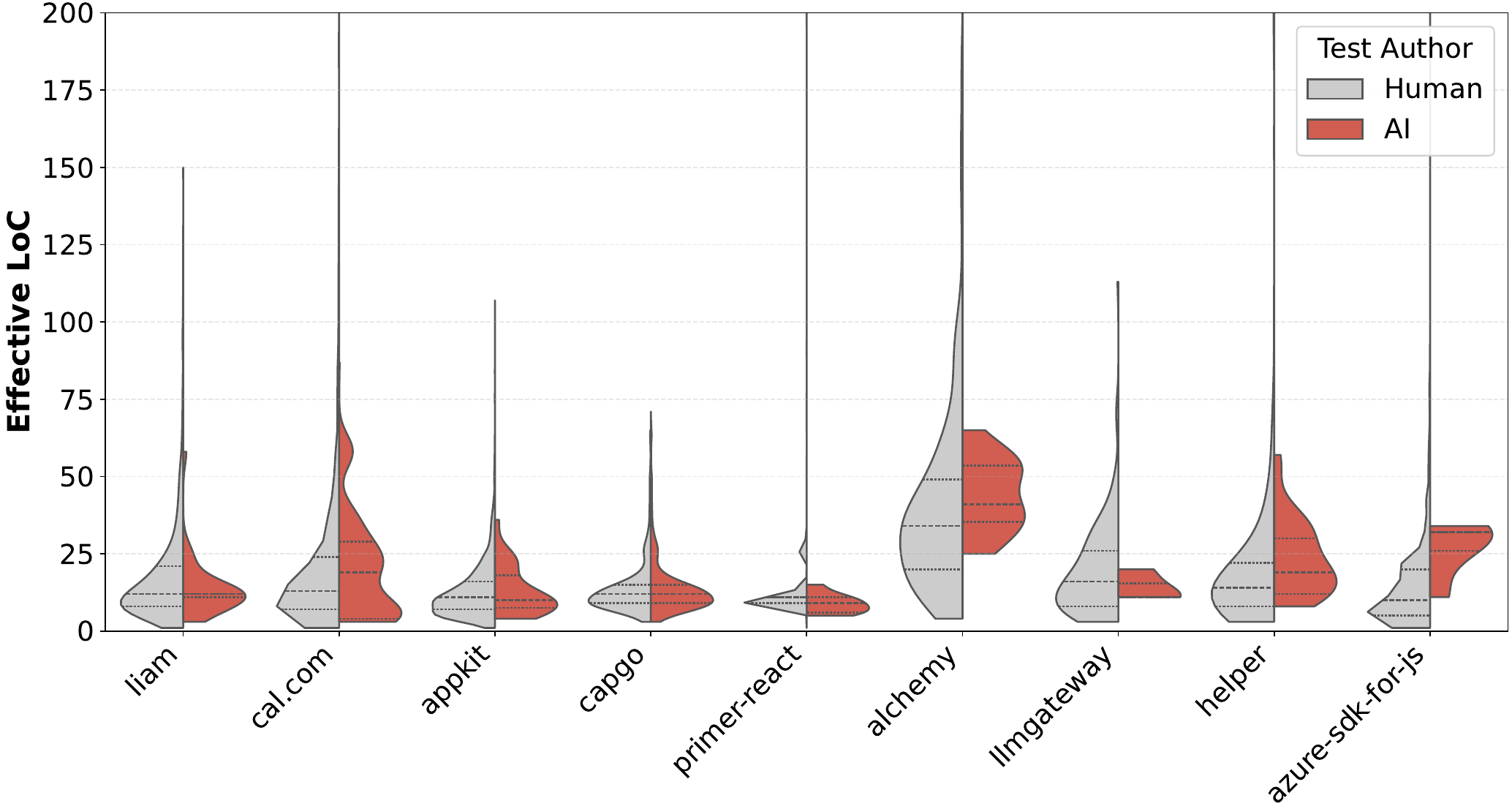}
        \caption{Effective eLoC Distribution}
        \label{fig:metric_loc}
    \end{subfigure}
    
    \vspace{-4pt} 
    
    \begin{subfigure}[b]{0.46\textwidth}
        \centering
        \includegraphics[width=0.8\linewidth]{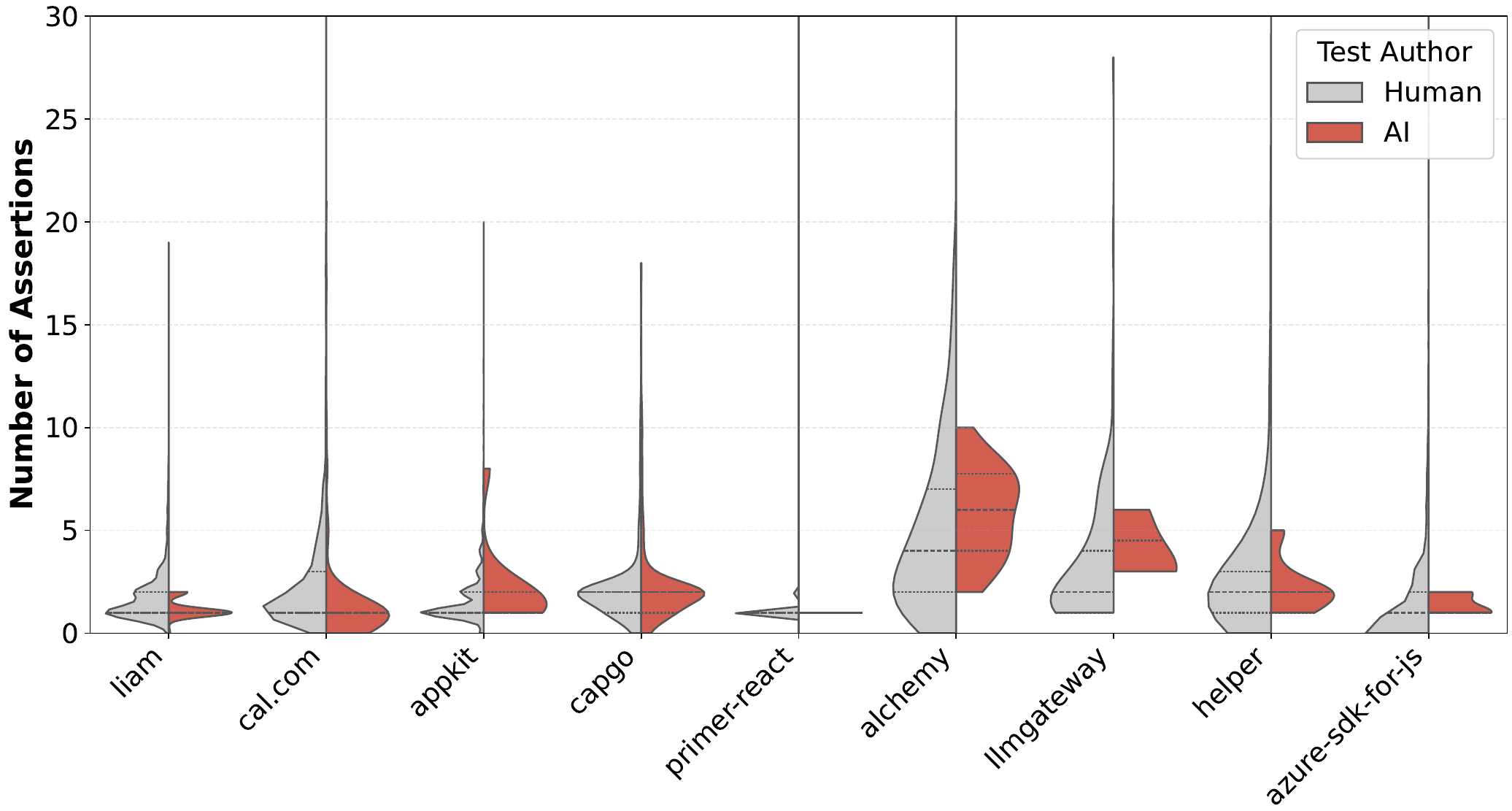}
        \caption{Number of Assertions Distribution}
        \label{fig:metric_assertion}
    \end{subfigure}


    \vspace{-4pt} 

    \setlength{\abovecaptionskip}{1pt}
    \caption{Comparison of Metrics: Human vs. AI Agents}
    \label{fig:metrics_distribution}
\end{figure}

\smallskip\noindent\textbf{Results.}
Figure~\ref{fig:metrics_distribution} presents the distribution of the three metrics for human-written and AI-generated tests. We used the Mann-Whitney U test~\cite{Mann1947OnAT} to assess statistical significance.
Of the 10 projects analyzed, 9 contained both AI-authored and human-authored test methods. \texttt{total-typescript-monorepo} was excluded because its AI-generated tests had been removed or modified by human developers in the latest release.

\smallskip\noindent\textbf{Effective eLoC: AI Tests are Longer and More Consistent.}
Our analysis reveals that AI-generated tests have a slightly higher median eLoC (12.0) compared to human-written tests (11.0). While the Mann-Whitney U test shows a statistically significant difference ($p < 0.001$), the calculated effect size is negligible (Cliff's $\delta = -0.12$), suggesting the central tendencies are practically similar.
However, a key observation lies in the variance and distribution (Figure~\ref{fig:metric_loc}). Human-written tests exhibit a long tail of outliers with high eLoC (Max: 699), whereas AI tests are confined to a narrower range (Max: 87). This suggests that while typical length is comparable, AI agents generate code with higher consistency, avoiding the ``massive'' test methods that occasionally occur in human-authored code.

\smallskip\noindent\textbf{Number of Assertions: More Verification Steps.}
AI-generated tests are characterized by a higher number of assertions (Figure~\ref{fig:metric_assertion}). The aggregated median is 2.00 for AI compared to 1.00 for humans ($p < 0.001$). This finding suggests that AI agents prioritize verbose setup followed by multiple verification steps within a single method. 

While this could be interpreted as thorough verification, it also suggests that AI agents may be prone to the \textit{Assertion Roulette} test smell---a known bad practice where multiple assertions in a single test make it difficult to pinpoint the exact cause of failure~\cite{vandeursen2001refactoring}. In contrast, human developers more frequently adhere to the ``one assertion per test'' principle, resulting in a lower median count. This structural difference highlights a potential trade-off: AI tests provide high verification density but may increase the maintenance effort required during debugging.

\smallskip\noindent\textbf{Cyclomatic Complexity: Simpler Logic with Fewer Outliers.}
Despite being longer, AI-generated tests do not exhibit higher complexity. Detailed complexity results are available in our online appendix~\cite{complexity}.
Significant differences were found in 2 out of 9 projects. The median complexity for both groups is 1.00, but the mean for AI (1.09) is lower than for humans (1.31). Crucially, AI tests rarely exceed a complexity of 2 or 3, lacking the high-complexity outliers seen in human tests. This indicates a strong preference by AI agents for linear, straightforward test logic without complex branching.





\begin{figure}[t]
    \centering
    \begin{subfigure}{0.9\linewidth} 
        \centering
        \includegraphics[width=\linewidth]{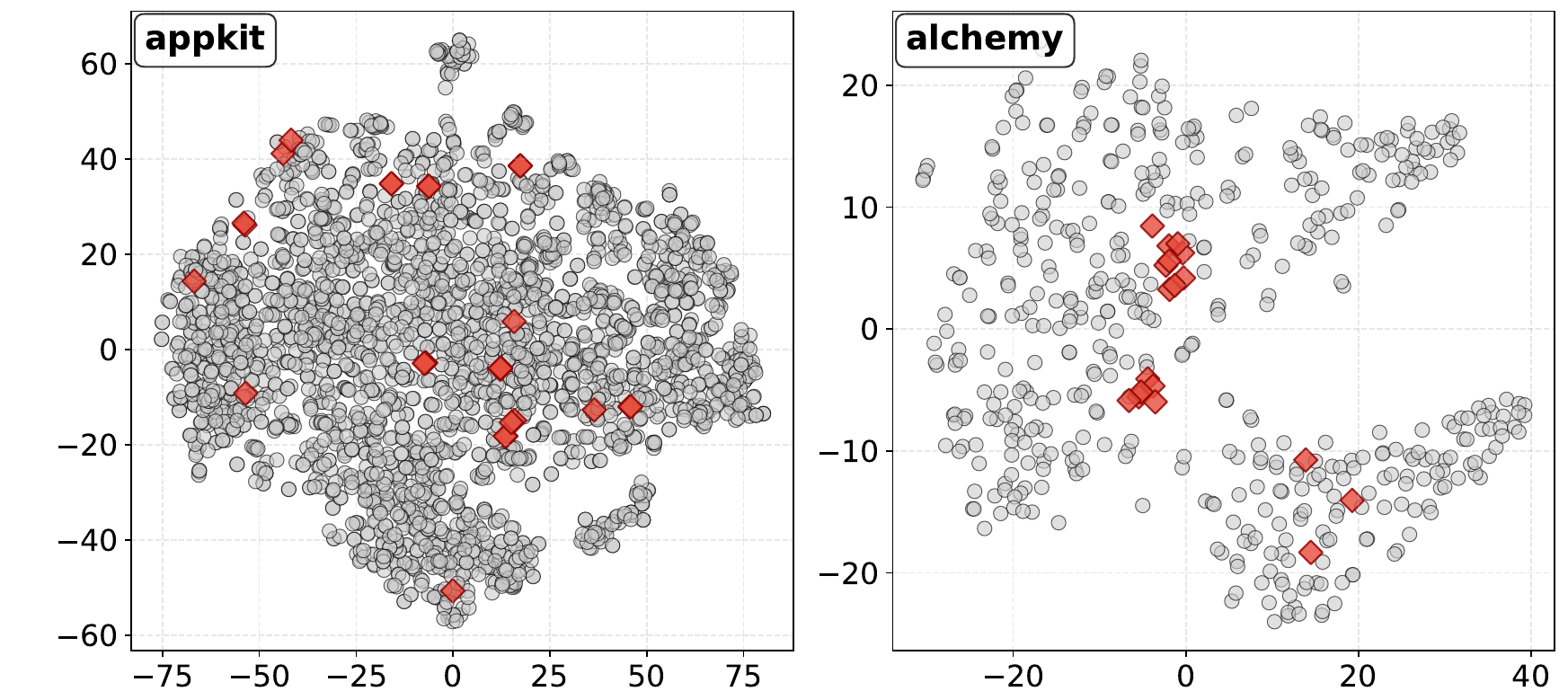}
        \vspace{-4mm} \\
        \makebox[0.5\linewidth][c]{\hspace*{7mm}\small (i) Dispersed Distribution}%
        \makebox[0.5\linewidth][c]{\hspace*{5mm}\small (ii) Localized Distribution}
    \end{subfigure}
    
    \begin{subfigure}{0.9\linewidth}
        \vspace{3mm}
        \centering
        \includegraphics[width=\linewidth]{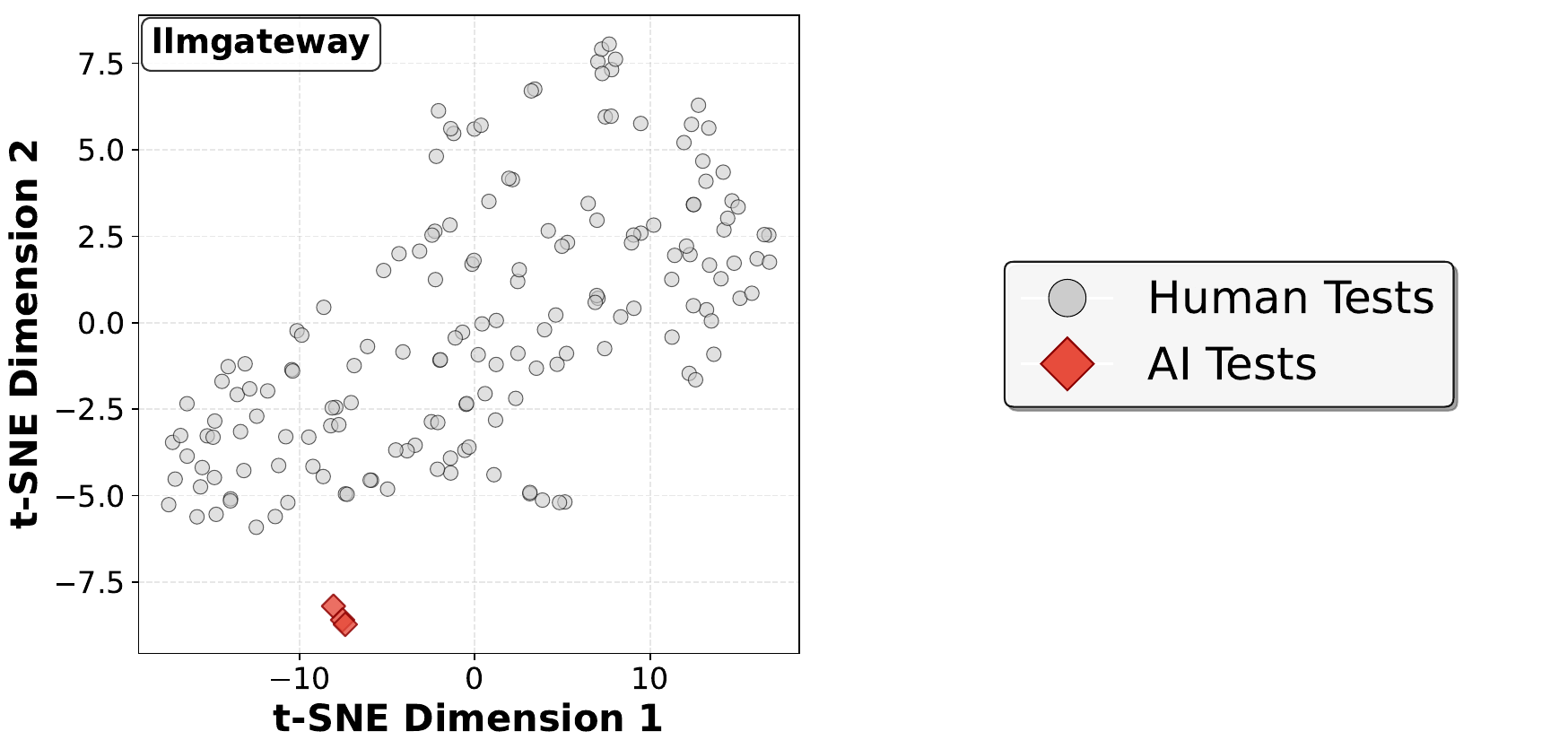}
        \vspace{-4mm} \\
        \makebox[0.5\linewidth][c]{\hspace*{7mm}\small (iii) Isolated Distribution}%
        \makebox[0.5\linewidth][c]{} 
    \end{subfigure}

    \vspace{-2mm}
    \caption{t-SNE Visualization of CodeBERT Embeddings.}
    \label{fig:tsne_combined}
    \vspace{4mm}
\end{figure}

\smallskip\noindent\textbf{Semantic Analysis with Embeddings.}
To capture characteristics beyond simple metrics, we visualized the semantic representations of the test methods using CodeBERT and t-SNE (Figure~\ref{fig:tsne_combined}).
By analyzing the distribution of AI-authored tests in the semantic space, we identified three distinct patterns of generation based on the spatial relationship between AI-generated and human-authored tests:
\textbf{(i) Dispersed Distribution}, where AI tests are broadly scattered throughout the entire human test distribution, indicating structural similarity to various human styles (\eg appkit, cal.com, and capgo); 
\textbf{(ii) Localized Distribution}, where AI tests overlap with only specific clusters or regions of human tests (\eg alchemy, liam, primer-react, helper); and 
\textbf{(iii) Isolated Distribution}, where AI tests are concentrated in limited regions that are spatially separated from most human tests (\eg llmgateway).
Visualization results for all projects are available in our replication package~\cite{ReplicationPackage}.

\begin{acmbox}

    \textbf{Answer to RQ2.}
    AI-generated tests tend to be longer with more assertions, yet exhibit lower or equal complexity, suggesting a preference for linear logic over complex branching. 
    
\end{acmbox}




\smallskip
\subsection*{\rqC}\label{sec:rqc}
\noindent\textbf{Motivation.} While adding tests is one thing, their practical value is often measured by their ability to exercise untested code~\cite{DBLP:conf/sigsoft/IvankovicPJF19}. We investigate the tangible impact of AI-generated tests by measuring the extent to which they contribute to the project's code coverage.

\smallskip\noindent\textbf{Approach.} To assess the impact of AI-generated tests on project quality, we measured coverage changes before and after test-adding commits. For each test-adding commit identified in RQ1, we calculated coverage at two points: the parent commit and the test-adding commit itself. Comparing these metrics allowed us to assess how AI-authored tests contributed to coverage improvement.

\smallskip\noindent\textbf{Results.}
Table~\ref{tab:rq3_coverage_results} summarizes the impact of AI-authored and human-authored test additions on code coverage. Of 10 projects analyzed in RQ1 and RQ2, we obtained coverage data for 531 test-adding commits across 3 projects where automated execution was feasible (other projects require project-specific setup).

In \texttt{liam} and \texttt{appkit}, AI-authored tests contributed positively to statement coverage, with average increases of +0.072 pts and +0.030 pts, respectively. In \texttt{liam}, among the 12 AI commits that changed statement coverage, 9 commits (75\%) resulted in improvement. In \texttt{appkit}, AI-authored tests showed more substantial improvement in branch coverage (+0.183 pts) compared to human-authored tests (+0.008 pts).
For \texttt{helper}, AI-authored tests showed nearly no changes in statement and branch coverage, while human contributions led to a +0.213 pts increase in branch coverage. 



\begin{table}[t]
\small
\centering
\caption{Average Coverage Change}
\label{tab:rq3_coverage_results}
\setlength{\tabcolsep}{6pt}
\begin{tabular}{llcc}
\toprule
Project & Author & Statement & Branch \\ \midrule
\texttt{liam} & AI ($N=18$) & +0.072 & +0.074 \\
 & Human ($N=143$) & -0.090 & -0.142 \\ \midrule
\texttt{appkit} & AI ($N=28$) & +0.030 & +0.183 \\
 & Human ($N=221$) & -0.001 & +0.008 \\ \midrule
\texttt{helper} & AI ($N=13$) & 0.000 & +0.006 \\
 & Human ($N=108$) & 0.000 & +0.213 \\ \bottomrule
\\
\end{tabular}
\end{table}

\vspace{1mm}

\begin{acmbox}

    \textbf{Answer to RQ3.}
    AI agents improve code coverage at levels comparable to or exceeding human developers. In \texttt{liam} and \texttt{appkit}, AI-generated tests yielded higher statement and branch coverage gains than humans. 
    
\end{acmbox}




\section{Related Work}\label{sec:related_work}

Automated test generation has evolved significantly, shifting from traditional Search-Based Software Testing to LLM-driven approaches \cite{DBLP:conf/icst/McMinn11, DBLP:journals/tse/SchaferNET24}.
Traditional tools like EvoSuite~\cite{DBLP:conf/sigsoft/FraserA11} excel at maximizing coverage but often produce unreadable code~\cite{DBLP:journals/ese/Arcuri18}. 
While recent LLM-based methods have improved readability~\cite{DBLP:conf/icse/DeljouyiKIZ25}, simple prompt-based approaches suffer from hallucinations and lack project-wide context~\cite{DBLP:conf/icse/000300L025}, frequently leading to non-compilable code~\cite{DBLP:journals/corr/abs-2305-04207}.

To address these limitations, recent research has focused on agent-based generation, demonstrating its superior performance through iterative refinement~\cite{DBLP:conf/nips/MundlerMHV24, DBLP:conf/ijcai/ChenDSZS00024, DBLP:conf/iclr/WangWAZZ25}. Studies indicate that agents employing self-correction loops significantly outperform single-turn LLMs~\cite{DBLP:conf/iclr/ChenLSZ24, DBLP:conf/fllm/RenzeG24a}. Furthermore, comparative evaluations reveal that Code Agents designed for repair can effectively generate reproducing tests and achieve high coverage~\cite{DBLP:conf/nips/YangJWLYNP24}.

However, these evaluations are largely limited to controlled benchmarks such as SWT-Bench~\cite{DBLP:conf/iclr/JimenezYWYPPN24}, leaving the prevalence and quality of agent-generated tests in the wild unexplored. To address this gap, we analyze the AIDev dataset~\cite{DBLP:journals/corr/abs-2507-15003}, examining agent-generated tests in actual development.



\section{Conclusion}\label{sec:conclusion}

This study presented an empirical analysis of test generation by AI agents using the AIDev dataset. Our investigation of \totalAnalyzedCommits test-related commits revealed that AI agents authored \totalAiRatioAmongTests of all test-adding commits, demonstrating their significant role in modern software development.
Structural analysis showed that AI-generated tests feature longer code and higher assertion density than human-written tests, while maintaining lower cyclomatic complexity through linear logic. AI-generated tests achieve code coverage gains comparable to human-authored ones, though their testing scope tends to be more localized in complex contributions.

Future research should fully evaluate long-term quality. First, \textit{mutation testing} should determine if the high assertion density of AI tests translates into superior fault-detection capabilities. Second, systematic investigation of \textit{test smells} (\eg Assertion Roulette) is essential to assess whether AI-generated patterns introduce technical debt that could hinder future maintenance.

\noindent
\textbf{Replication Packages:} We provide the program and data used in our replication package~\cite{ReplicationPackage}.


\vspace{-1mm}
\begin{acks}
This work was supported by JSPS KAKENHI (JP24K02921, JP25K21359) and JST PRESTO (JPMJPR22P3), ASPIRE (JPMJAP2415), and AIP Accelerated Program (JPMJCR25U7).
\end{acks}

\balance
\bibliographystyle{ACM-Reference-Format}
\bibliography{references}


\end{document}